\documentclass[12pt,a4paper]{iopart}

\usepackage[pdftex]{hyperref}
\usepackage{xcolor}

\usepackage{float}
\restylefloat{table}

\usepackage{graphicx}
\usepackage{amssymb}

\usepackage{longtable}
\usepackage{rotating}
\usepackage{color}
\usepackage{epsf}
\usepackage{fancyhdr}
\usepackage{ifthen}
\usepackage{slashed}

\usepackage{placeins}

\begin{document}

 \title[Minimally-modeled search of higher multipole gravitational radiation]{Minimally-modeled search of higher multipole gravitational-wave radiation in compact binary coalescences.}

 
\author{G.~Vedovato$^{1}$, E.~Milotti$^{2,3}$, G.~A.~Prodi$^{4,5}$, S.~Bini$^{6,5}$, M.~Drago$^{7,8}$, V.~Gayathri$^{9}$, O.~Halim$^{3}$, C.~Lazzaro$^{10,1}$, D.~Lopez$^{11}$, A.~Miani$^{6,5}$, {B.~O'Brian$^{9}$}, F.~Salemi$^{6,5}$, M.~Szczepanczyk$^{9}$, S.~Tiwari$^{11}$, A.~Virtuoso$^{2,3}$, S.~Klimenko$^{9}$}
\address{$^{1}$ INFN, Sezione di Padova, I-35131 Padova, Italy }
\address{$^{2}$ Universit\`a di Trieste, Dipartimento di Fisica, I-34127 Trieste, Italy }
\address{$^{3}$ INFN Sezione di Trieste, I-34127 Trieste, Italy }
\address{$^{4}$ Universit\`a di Trento, Dipartimento di Matematica, I-38123 Povo, Trento, Italy}%
\address{$^{5}$ INFN, TIFPA, I-38123 Povo, Trento, Italy}%
\address{$^{6}$ Universit\`a di Trento, Dipartimento di Fisica, I-38123 Povo, Trento, Italy}%
\address{$^{7}$ Universit\`a di Roma  La Sapienza, I-00185 Roma, Italy}
\address{$^{8}$ INFN, Sezione di Roma, I-00185 Roma, Italy}%
\address{$^{9}$ University of Florida, Gainesville, FL 32611, USA }%
\address{$^{10}$ Universit\`a di Padova, Dipartimento di Fisica e Astronomia, I-35131 Padova, Italy }
\address{$^{11}$ Physik-Institut, University of Zurich, Winterthurerstrasse 190, 8057 Zurich, Switzerland }
\ead{claudia.lazzaro@pd.infn.it}

\date{\today}

\begin{abstract}

As the Advanced LIGO and Advanced Virgo interferometers, soon to be joined by the KAGRA interferometer, increase their sensitivity, they detect an ever-larger number of gravitational waves with a significant presence of higher multipoles in addition to the dominant $(2, 2)$ multipole. These higher multipoles can be detected with different approaches, such as the minimally-modeled burst search methods, and here we discuss one such approach based on the coherent WaveBurst pipeline (cWB). 
During the inspiral phase the higher multipoles produce chirps whose instantaneous frequency is a multiple of the dominant $(2, 2)$ multipole, and here we describe how cWB can be used to detect these spectral features. The search is performed within suitable regions of the time-frequency representation; their shape is determined by optimizing the Receiver Operating Characteristics.
This novel method has already been used in the GW190814 discovery paper (Astrophys. J. Lett. \textbf{896} L44) and is very fast and flexible. Here we describe in full detail the procedure used to detect the $(3,3)$ multipole in GW190814 as well as searches for other higher multipoles during the inspiral phase, and apply it to another event that displays higher multipoles, GW190412, replicating the results obtained with different methods. 
The procedure described here can be used for the fast analysis of higher multipoles and to support the findings obtained with the model-based Bayesian parameter estimates. 
\end{abstract}

\submitto{\CQG}

\maketitle

\pagebreak








\section{Introduction}

The first three observing runs of the Advanced LIGO and Advanced Virgo interferometers have produced a wealth of results on coalescing binary systems \cite{LIGOScientific:2018mvr,Abbott:2020niy}. Now, as the interferometers of the LIGO-Virgo-KAGRA (LVK) network gradually increase their sensitivity \cite{Aasi:2013wyav10,LIGOScientific:2014pky,VIRGO:2014yos,KAGRA:2020tym}, they probe more deeply into the population of coalescing binary black hole systems, potentially displaying features such as  asymmetric masses, higher inclination, precession, and orbital eccentricity. When these features are present, in addition to the ever-present quadrupole gravitational radiation, a non-negligible fraction of the energy radiated as gravitational waves is carried by the higher multipoles (HM) \cite{Blanchet:2013haa, Mills:2020thr,Divyajyoti:2021uty}. The analysis of recent observations of compact binary coalescences (CBC) made by the LIGO-Virgo Collaboration demonstrated the existence of HMs in detected signals \cite{LIGOScientific:2020stg,GW190521Adiscovery,Abbott:2020khf,Abbott:2020niy}. The HMs carry important information on the coalescing binary system, however a description of the signals that includes HMs requires more sophisticated waveform models \cite{Khan:2018fmp,Khan:2019kot,Ossokine:2020kjp,Babak:2016tgq,Varma:2019csw,London:2017bcn,Cotesta:2018fcv}. These waveform models are used in match filtered pipelines for the detection and reconstruction of gravitational wave signals with HMs \cite{Abbott:2020niy,LIGOScientific:2020stg,GW190521Adiscovery,Abbott:2020khf,Abbott:2020mjq}. However, HMs can also be detected by minimally-modeled burst algorithms, such as \textit{coherent WaveBurst} (cWB), a data analysis pipeline which is used both to detect and to reconstruct transient gravitational waves \cite{Klimenko_2016, Drago:2020kic,CalderonBustillo:2017skv}.  

\medskip

cWB identifies HMs in the time-frequency representation of gravitational-wave signals from compact binary coalescences by finding coherent excess power in chirp-like regions that correspond to different HMs. 
This method is similar -- although not equivalent -- to an alternative one described in ref. \cite{Roy:2019phx}, and it can be placed in the more general context of the procedures used to compare the cWB reconstructions with the
estimates obtained from Bayesian inference \cite{PhysRevD.100.042003}. 

\medskip

The paper is organized as follows:
section \ref{sec:waveform} gives a broad outline of the procedure and the definition of the test statistic, i.e. the waveform residual energy. 
Section \ref{sec:TFslice} specifies the choice of the relevant time-frequency region where the test statistic is evaluated; it also discusses the parameterization in terms of harmonics of the dominant quadrupole emission and the optimization by means of the Receiver Operating Characteristics.
Section \ref{sec:performances} focuses on the practical implementation of optimization for the gravitational waves GW190814 and GW190412. These two gravitational waves are relevant because they are the first ones to show unequivocal indications of HM presence, thanks to the large mass ratio of the components of these compact binary systems. 
Finally we discuss results and future perspectives in section \ref{sec:remarks}.

\section{Highlighting the higher multipoles}
\label{sec:waveform}

In cWB \cite{Klimenko_2016,Drago:2020kic}, waveforms are reconstructed by first decomposing the gravitational-wave signals with the discrete Wilson--Daubechies--Meyer (WDM) wavelet transform \cite{Necula:2012zz} to produce a time-frequency representation. The time-frequency pixels that represent individual wavelets in this representation are selected by retaining only a fixed fraction of them with larger excess network energy. Next, cWB estimates the coherent response of the gravitational-wave observatories, and separates it from the incoherent contribution of each detector, by means of the maximization of the constrained likelihood described in \cite{Klimenko_2016}. Finally, the coherent wavelets provide a reconstruction (a point estimate in the time domain) of the gravitational waveform (for a more complete overview of cWB, see \cite{Klimenko_2016,Drago:2020kic}).

\medskip

In \cite{PhysRevD.100.042003} we considered generic strategies to compare these waveform reconstructions with the estimates obtained by Bayesian inference methods based on detailed waveform models of compact binary coalescences; procedures like those described in \cite{PhysRevD.100.042003} have been used to test general relativity~\cite{TheLIGOScientific:2016src, LIGOScientific:2019fpa,Abbott:2020jks}, at least within the precision with which the method matches the predictions of Einstein's theory. They can also be used to detect signal features that are not described by the waveform models, like echos in the ringdown phase. 

\medskip

Here we describe a similar procedure specifically aimed at detecting HMs, where we use two waveform models that are very similar, except in their HM content (one includes HMs while the other does not). 

In general, all the waveforms we consider are whitened by the cWB pipeline as in \cite{PhysRevD.100.042003}, ``on-source data'' indicates data at the time of the gravitational wave, and ``off-source data'' indicates data at times that do not include gravitational wave detections and provide independent noise instances of the data. The use of off-source data at different times is necessary to assess the effect of actual noise fluctuations with no assumptions on noise statistics except wide-sense stationarity.  

Here, we compare the cWB reconstructions with model waveforms (both with and without HMs) obtained by the Bayesian inference methods \cite{Veitch_2015, Ashton_2019}, and to this end, we  define the following statistic, called  {\it waveform residual energy}, $E_\mathrm{res}$:
\begin{equation}
\label{Eres}
 E_\mathrm{res} =  \sum_{k=1}^\mathrm{det} \; \sum_{i\in\langle \mathrm{pixels} \rangle} (w_{k}^\mathrm{cWB}[i] - w_{k}^\mathrm{model}[i])^2,
\end{equation}
where  $w_{k}^\mathrm{cWB}[i]$ and $w_{k}^\mathrm{model}[i]$ are respectively the WDM transforms of the cWB reconstruction 
and of a waveform model, the index $k$ runs over all detectors, and the index $i$ runs over a specific subset of the WDM pixels (denoted here by the notation $\langle \mathrm{pixels} \rangle$) in the time-frequency representation (for more details see Sec. III.A of ref. \cite{PhysRevD.100.042003}).

\medskip

The consistency of each cWB point estimate with Bayesian estimation is measured by the residual energy $E_\mathrm{res}^\mathrm{(on-source)}$ between the cWB on-source reconstruction and the maximum likelihood (MaxL) sample waveform from the Bayesian analysis without HMs\footnote{Here we take the MaxL sample as the best estimate from the Bayesian parameter estimate procedure instead of the alternative \textit{maximum a posteriori} sample (MAP), which would be better motivated in a Bayesian perspective, because the priors are flat on specific regions in the parameter space, so that the MAP sample offers no real advantage when it happens to be well within the interior of these regions, and it may produce worse results if it rails against their boundaries.}, and its significance is evaluated by the empirical distribution obtained by off-source injections of random samples  from the posterior distribution into a wide off-source time interval at equally spaced times. 
Signals are injected off-source, they are reconstructed by cWB and compared with their whitened version without HM, to evaluate the residual energy. The set of residual energies defines an empirical distribution: using this distribution and the on-source residual energy $E_\mathrm{res}^\mathrm{(on-source)}$, we compute a corresponding \textit{p}-value, and thus put to the test the hypothesis that the injected waveform is in  good agreement with the cWB reconstruction.

\medskip

As an example, consider the left panel in Figure \ref{f:TFslice}: the plot displays the pixel-by-pixel $ E_\mathrm{res}$ for all the pixels selected in the cWB on-source analysis of GW190814, where the reference waveform is the MaxL estimate obtained with the {\tt SEOBNRv4\_ROM} model \cite{Bohe:2016gbl} which does not include HMs {(here and throughout the paper we used the same data from the LIGO Hanford, LIGO Livingston and Virgo interferometers as in the GW190814 discovery paper \cite{Abbott:2020khf})}. The plot has the characteristic chirp-like shape found in CBCs, but it appears to be wider than usual: as we shall see in the following, this corresponds to a significant deviation of $E_\mathrm{res}$ from the null hypothesis, and to the demonstrable presence of the $(3,3)$ multipole.

\medskip

Indeed, the instantaneous frequency of the generic $(\ell, m)$ multipole\footnote{We use  $(\ell, m)$ as a shorthand for both $(\ell, m)$ and $(\ell, -m)$.} emitted by spinning, non-precessing black hole binaries,  is to a good approximation a scaled version of the dominant $(2, 2)$ multipole \cite{Blanchet:2013haa,London:2017bcn}
\begin{equation}
\label{fmodes}
f_{\ell,m}(t) \approx \frac{m}{2} f_{22}(t).
\end{equation}
Therefore, in comparisons between the cWB reconstructions and the MaxL waveforms obtained with  models that do not include HMs, we look for the presence of significant residual energy {by integrating the pixel residual energy} along ``slices'' of the time-frequency map; a slice is defined by the region between the curves  $f(t) = (\alpha\pm \delta \alpha) f_{22}(t)$, where  $\alpha$ is a non-negative real parameter~\cite{LIGOScientific:2020stg,Roy:2019phx} and $\delta\alpha$ determines the strip width, between a minimum and a maximum time (see the right panel of Figure \ref{f:TFslice} for an example). 

\medskip 

\begin{figure*}[ht!]
\begin{centering}
\includegraphics[width=0.6\textwidth]{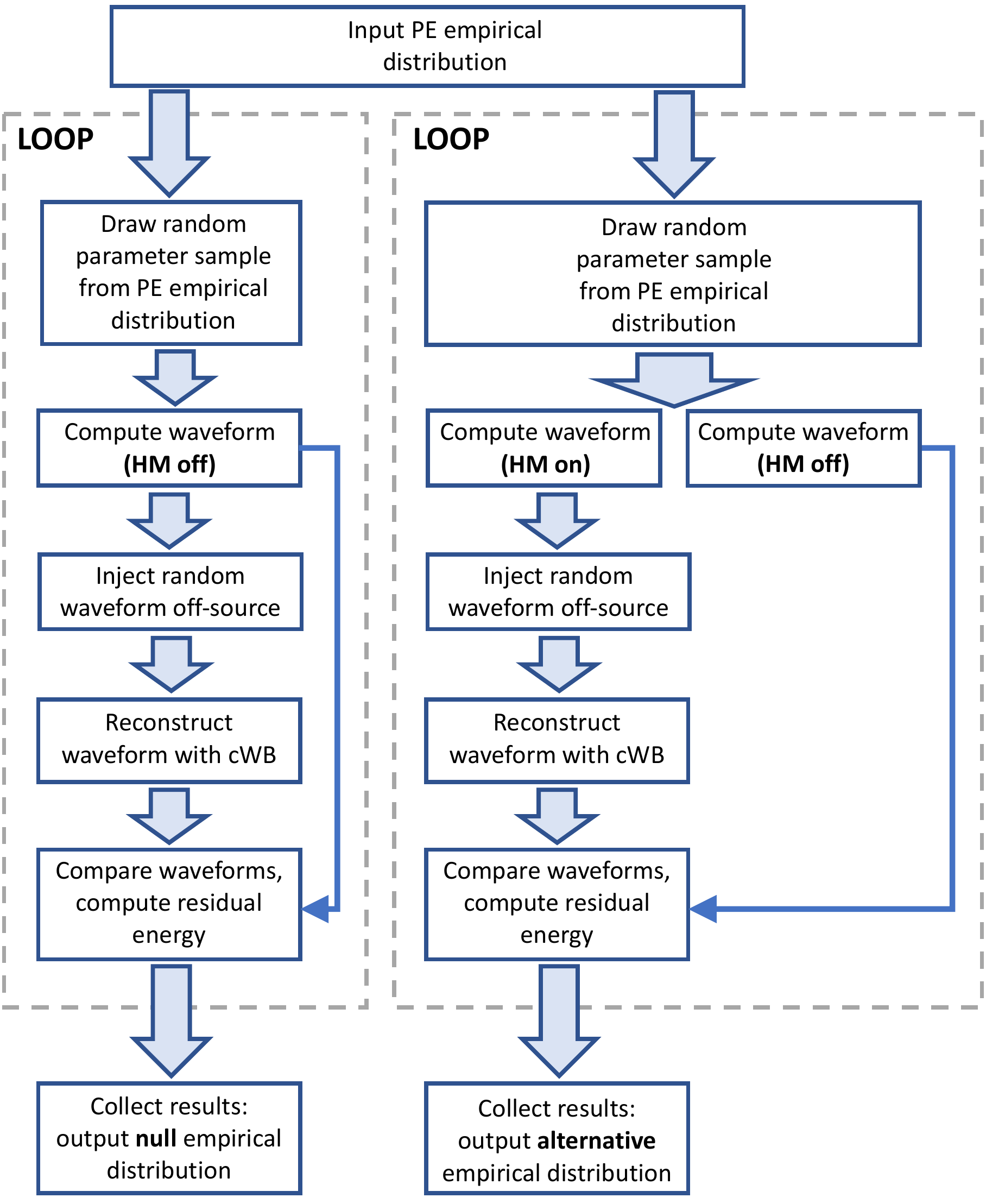}
\end{centering}	
\caption{\label{f:FlowChart} {Schematic structure of the Monte Carlo program used to produce the  null and alternative empirical distributions. We remark that the two different cWB reconstructions are both compared with the waveform without HMs (this is indicated by the solid blue arrows). Typically, the Monte Carlo loop is repeated over about 2000 off-source injections. } }
\end{figure*}

\begin{figure*}[ht!]
\includegraphics[clip,height=0.28\textheight]{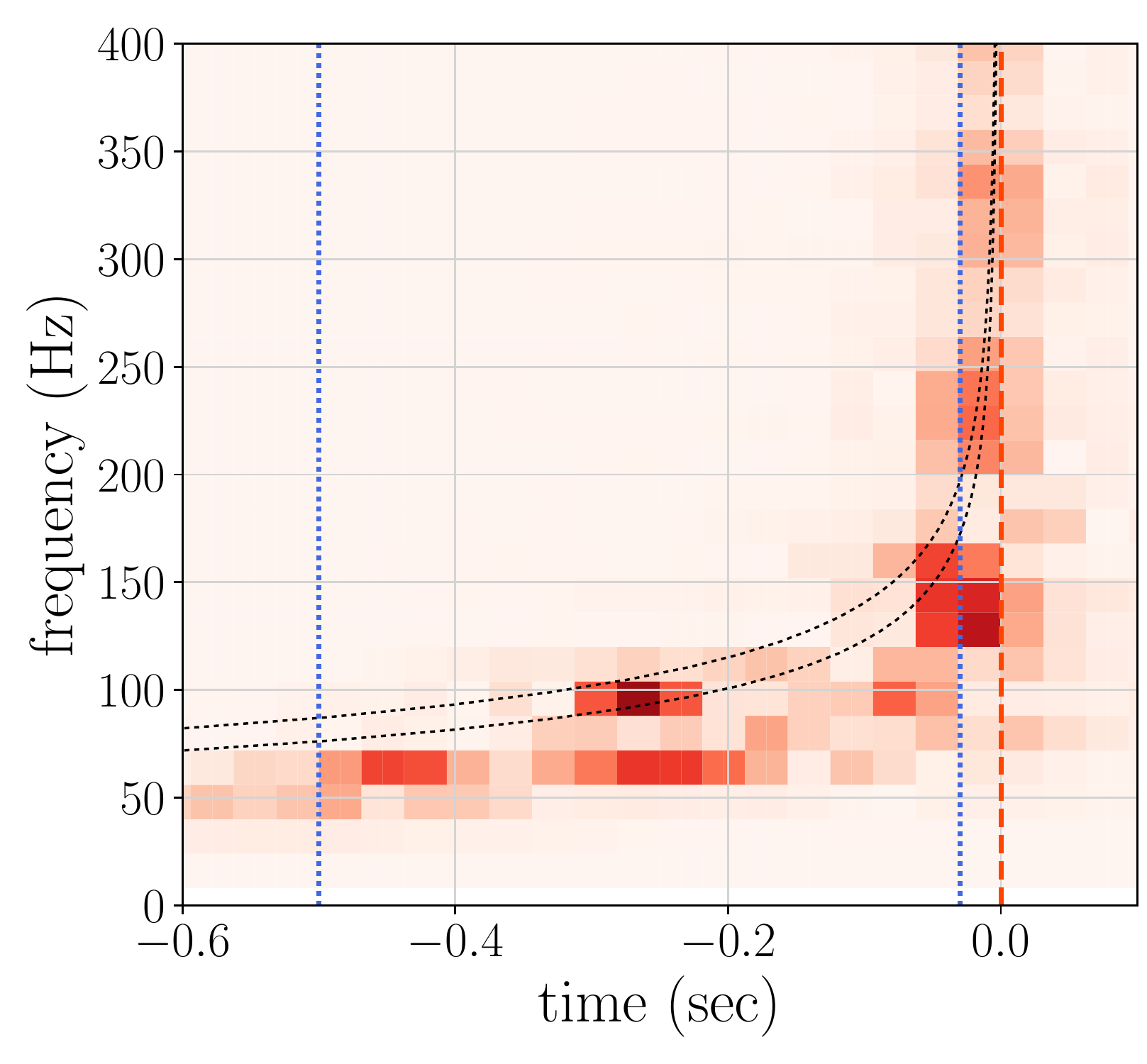} \hspace{0.02\textwidth}
\includegraphics[clip,height=0.28\textheight]{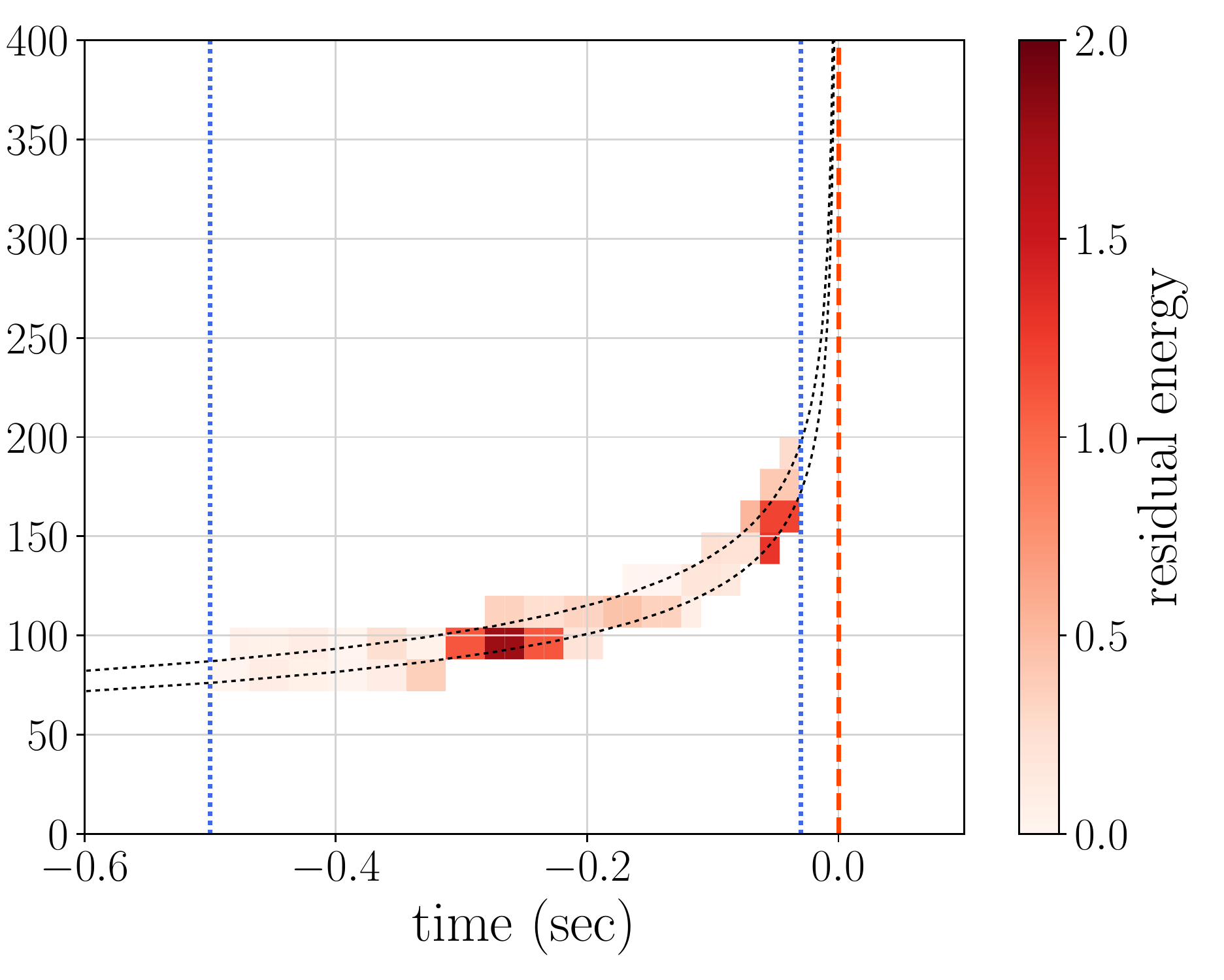} \\
\caption{\label{f:TFslice} Left panel: time-frequency representation of the pixel-by-pixel residual energy of the GW190814 event obtained by cWB with respect to the MaxL {\tt SEOBNRv4\_ROM} waveform, projected into the LIGO Livingston detector using the WDM transform with resolution $dt = 1/32$~s and $df = 16$~Hz. Right panel: same map, showing only the time-frequency pixels which overlap at least partially the time-frequency slice with $\alpha=1.5$ (corresponding to the  $(3,3)$ HM): these pixels are used to evaluate the total residual energy $E_\mathrm{res}(\alpha; \delta \alpha, \Delta t, \delta t, df)$ in the time-frequency slice. The red vertical line shows the merger time  (GPS time: 1249852257.0154 s) from the MaxL {\tt SEOBNRv4\_ROM} waveform determined by Bayesian inference, referred to the LIGO Livingston detector. The dotted blue vertical lines show the selected time range of the time-frequency slice,  $[t_\mathrm{merger} - 0.5$~s$, t_\mathrm{merger} - 0.03$~s$]$. The black dotted curves show the boundaries of the time-frequency slice, defined by $[\alpha- 0.1, \alpha+ 0.1] \times f_{22}(t)$ with $\alpha = 1.5$. The fluctuations of $E_\mathrm{res}$ in pixels corresponding to the $(2,2)$ multipole in the left panel may appear large but they are compatible with the null hypothesis for this multipole (as shown later, see figure \ref{f:GW190814-pvalue}).}
\end{figure*}

We establish the statistical significance of the excess residual energy \emph{in each time-frequency slice} by carrying out  Monte Carlo simulations: 
we inject random waveform samples from the posterior distribution obtained from Bayesian inference into off-source data, both with a model that does not include HMs and with one that instead does include them, and in each case we compute the residual energy between the cWB reconstruction and the model waveform without HMs. Thus, we obtain two empirical distributions of residual energy, one for the null hypothesis (no HMs) and one for the alternative hypothesis. 
{This scheme is graphically illustrated in figure \ref{f:FlowChart}}.

Thus, we obtain two empirical distributions of residual energy, one for the null hypothesis (no HMs) and one for the alternative hypothesis. We remark that the resulting cWB reconstructions include statistical fluctuations both from Bayesian inference and from the variable off-source background noise. Finally, using these empirical distributions we compute the \textit{p}-values corresponding to the $E_\mathrm{res}^\mathrm{(on-source)}$ computed in each slice with the model with no HMs, for both hypotheses and for all time-frequency slices. 

If the slice contains significant residual energy from a HM (see, e.g., the slice in the right panel of Figure \ref{f:TFslice}), then the \textit{p}-value for the null hypothesis is small, while the \textit{p}-value for the alternative hypothesis is significantly larger.

\section{Tuning the shape of the time-frequency slices}
\label{sec:TFslice}

Different slice widths $2\delta\alpha $ produce different results: narrow slices (small $\delta \alpha$) are greatly affected by background noise, while wide slices (large $\delta \alpha$) have low $\alpha$-resolution. In this section, we describe the tuning of $\delta\alpha$ and of the other parameters that determine both the false alarm probability (the \textit{p}-value of the null hypothesis) and the detection efficiency (the \textit{p}-value of the alternative hypothesis). To this end, we optimize the procedure targeting the strongest among the HMs, i.e, the $(3,3)$ multipole ($\alpha=1.5$), using a well-established method, the Receiver Operating Characteristic (ROC) \cite{FAWCETT2006861}, to strike the best balance between detection efficiency (defined as the \textit{p}-value of the alternative hypothesis) and false alarm probability (the \textit{p}-value of the null hypothesis).

\medskip 

We start by selecting a time interval  $(t_\mathrm{merger} - \Delta t, t_\mathrm{merger} - \delta t)$ which spans the late inspiral phase to maximize the chance of observing the effects of HMs in the time-frequency slices determined by equation (\ref{fmodes}). The lower bound  $t_\mathrm{merger} - \Delta t$, where $\Delta t$ is roughly the duration of the late inspiral phase, which is visible in the highest-sensitivity frequency band of the detectors and is motivated by the robust expectation that HM emission is strongest  during late inspiral; the upper bound $t_\mathrm{merger} - \delta t$, where the interval $\delta t$  is of the order of the pixel time resolution,  is used to avoid  merger effects \cite{Blanchet:2013haa} that may leak into the inspiral phase because of the pixel time resolution.

Having completed the definition of the time-frequency slices, we scan over the $\alpha$ range $0.25 - \delta \alpha \le \alpha \le 2.75 + \delta \alpha$  (this covers $1\le m \le 5$) with step size $\leq \delta \alpha$ (see Figure \ref{f:TFslice}). The procedure returns multiple residual energy estimates which are correlated because of the partial overlap of the time-frequency slices. 

This residual energy $E_\mathrm{res}(\alpha; \delta \alpha, \Delta t, \delta t, df)$ is obtained by summing over the time-frequency pixels which belong to the time-frequency slice determined by $\alpha$ and $\delta \alpha$. The sum includes the contribution from those 
pixels which belong only partly to the slice by weighting their residual energy by the fraction of pixel area inside the slice: this mitigates effects due to the finite time-frequency resolution of the WDM representation, but also adds some correlation in the residual energy of adjacent time-frequency slices. The time-frequency resolution of the WDM wavelet representation (determined by the frequency step $df$ for a fixed time-frequency pixel area) is also a free parameter of the procedure and must be optimized along with the other parameters. 

\section{The cases of GW190814 and GW190412}
\label{sec:performances}
During the third observing run (O3), the Advanced LIGO and Advanced Virgo interferometers observed two events with very asymmetric masses, GW190814 ($q=0.112^{+0.008}_{-0.009}$, \cite{Abbott:2020khf}) and GW190412 ($q=0.28^{+0.12}_{-0.07}$, \cite{LIGOScientific:2020stg}), where the presence of HMs was detected using different analysis methods. 
In the following, we describe the parameter optimization procedure with the ROC curves in the specific case of the GW190814 gravitational wave.

\begin{figure*}[h!]
\begin{center}
\includegraphics[width=0.8\textwidth]{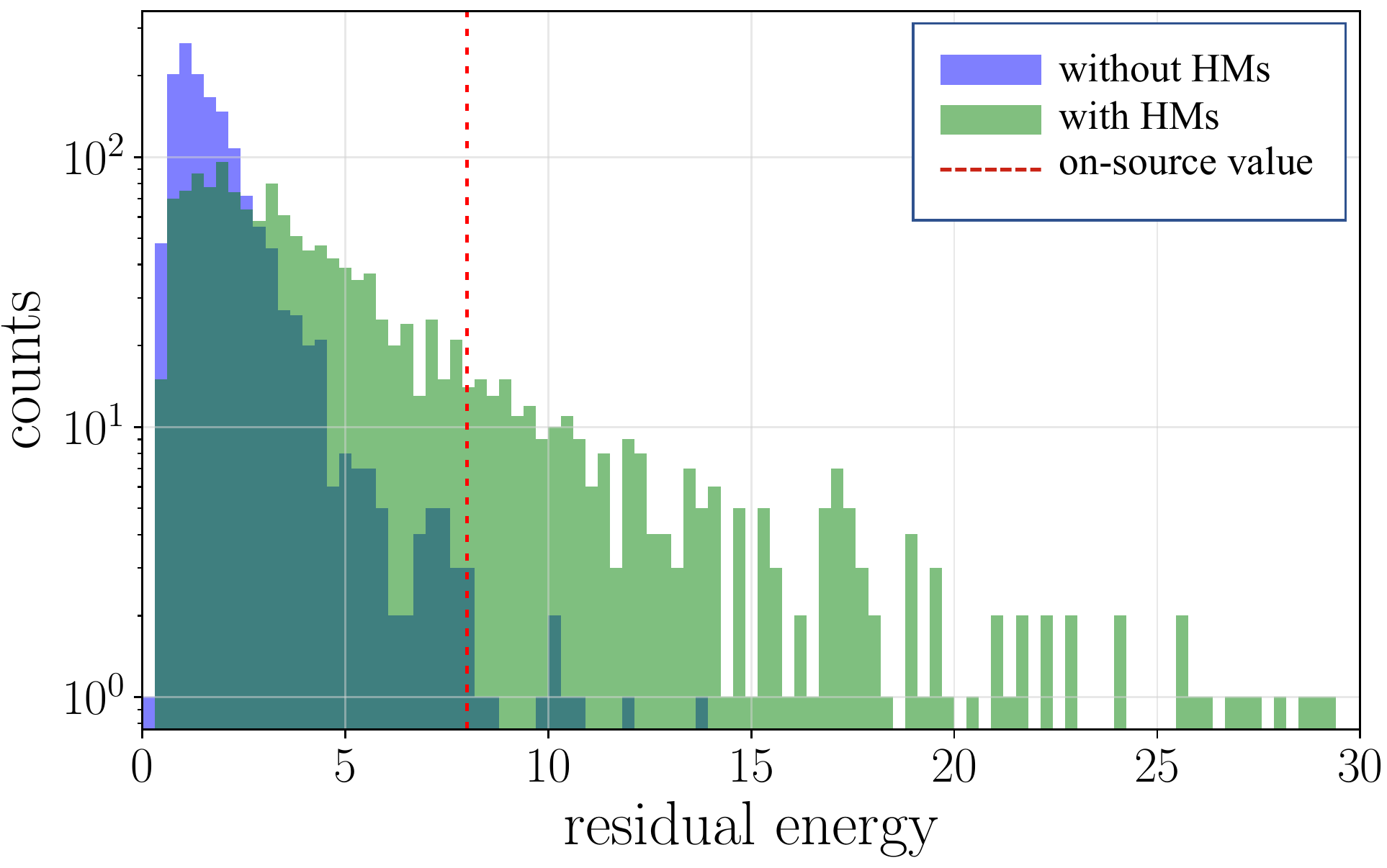}
\end{center}	
\caption{\label{f:distributions} Empirical residual energy distributions for the time-frequency slice defined by $\alpha = 1.5$, $\delta \alpha =0.1$, $\Delta t = 0.5$~s, and $\delta t = 0.03$~s in our study of the GW190814 HMs. Red vertical line: on-source result for GW190814. Purple histogram: empirical distribution for the null hypothesis, from {\tt SEOBNRv4\_ROM} injections in off-source data. Green histogram: empirical distribution for the model with higher order modes, from {\tt SEOBNRv4HM\_ROM} injections in off-source data. The GW190814 on-source residual energy with respect to the MaxL waveform for model with HMs (red dashed line) is an outlier of the null model, {\tt SEOBNRv4\_ROM} with its \textit{p}-value~=~0.0068, but it is compatible with {\tt SEOBNRv4HM\_ROM}, the waveform model that includes higher multipoles (\textit{p}-value~=~0.17).}
\end{figure*}

We used two waveform models: a Spin-aligned Effective One Body waveform model without higher multipoles, {\tt SEOBNRv4\_ROM} \cite{Bohe:2016gbl} as implemented in LALSuite \cite{lalsuite} (LALSim version 1.10.0.1), to estimate the false alarm probability, and a version of the same model that includes the higher multipoles (2,1), (3,3), (4,4), and (5,5) along with the dominant (2,2) multipole,  {\tt SEOBNRv4HM\_ROM} \cite{Cotesta:2018fcv,lalsuite,Cotesta:2020qhw}, to estimate the detection efficiency. The  {\tt SEOBNRv4HM\_ROM} model reduces to {\tt SEOBNRv4\_ROM} by turning off the HMs\footnote{
{the {\tt SEOBNRv4} model has been chosen because it exists in two versions, with and without HMs; moreover, within the LALSuite implementation of {\tt SEOBNRv4HM\_ROM} it is even possible to select the generation of a single multipole, like, e.g., (3,3).}}. 
Figure \ref{f:distributions} shows an example of the empirical distributions that we obtain in our $\alpha$ scan, for the  slice defined by $\alpha = 1.5$, $\delta \alpha =0.1$, $\Delta t = 0.5$~s, and $\delta t = 0.03$~s; from these empirical distributions and from the on-source result we find that  the GW190814 on-source residual energy with respect to the MaxL waveform for model with HMs is an outlier of the null model, {\tt SEOBNRv4\_ROM} with its \textit{p}-value~=~0.0068, but it is compatible with {\tt SEOBNRv4HM\_ROM}, the waveform model that includes higher multipoles (\textit{p}-value~=~0.17).

\begin{figure*}[ht!]
\begin{center}
\includegraphics[clip,width=0.8\textwidth]{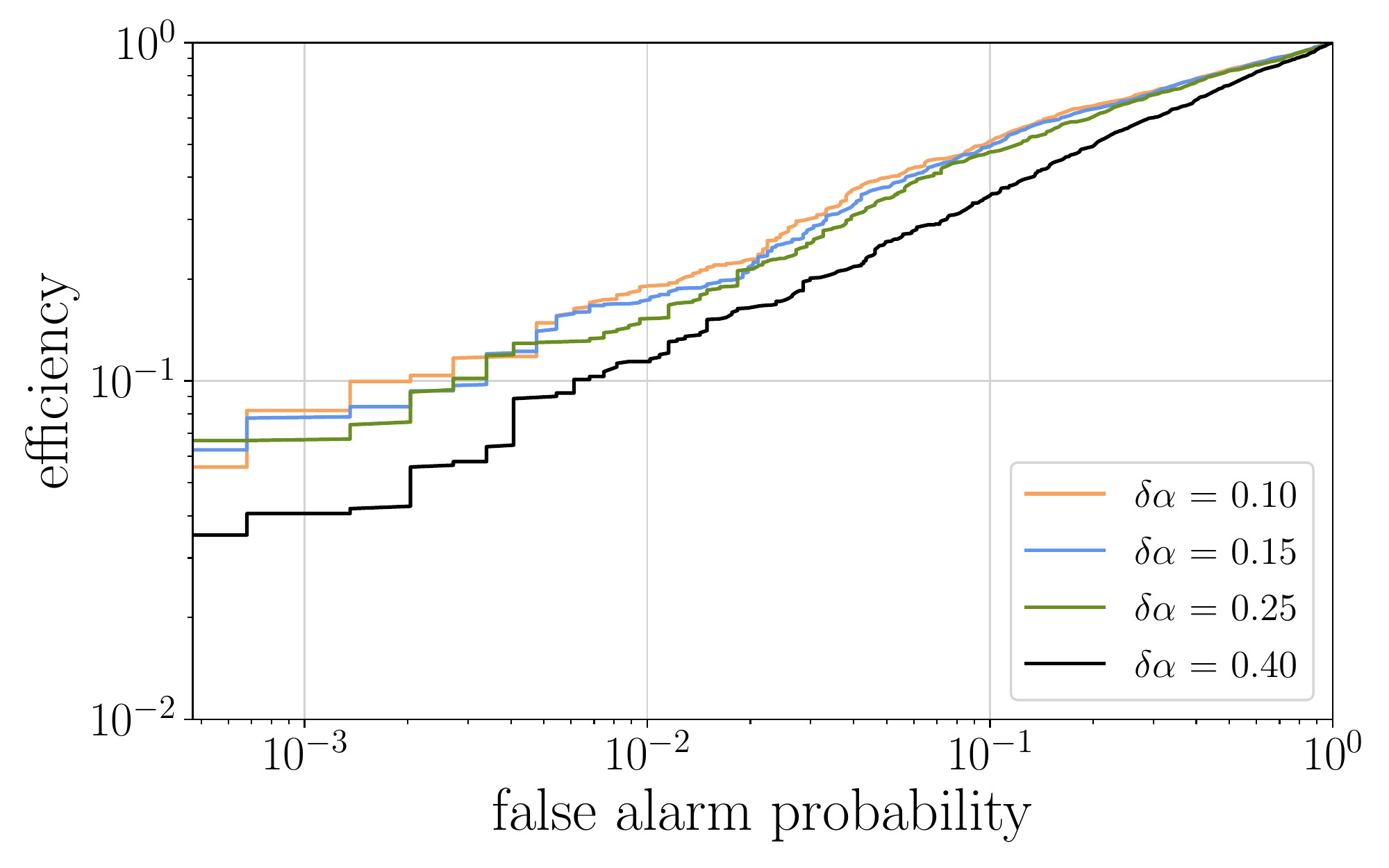}\hfil
\end{center}	
	\caption{\label{f:ROC} 
	Receiver Operating Characteristic curves (detection efficiency vs. false alarm probability) for $E_\mathrm{res}(\alpha; \delta \alpha, \Delta t, \delta t, df)$, for different choices of the $\delta \alpha$ parameter, with $df=16$~Hz, $\Delta t = 0.5$~s, $\delta t =0$ {in the case of GW190814}. The best value of $\delta \alpha$ corresponds to the uppermost curve, because it has almost everywhere the highest detection efficiency for a given false alarm probability (\textit{p}-value).}
\end{figure*}

\begin{figure*}[h!]
\begin{center}
\includegraphics[width=0.8\textwidth]{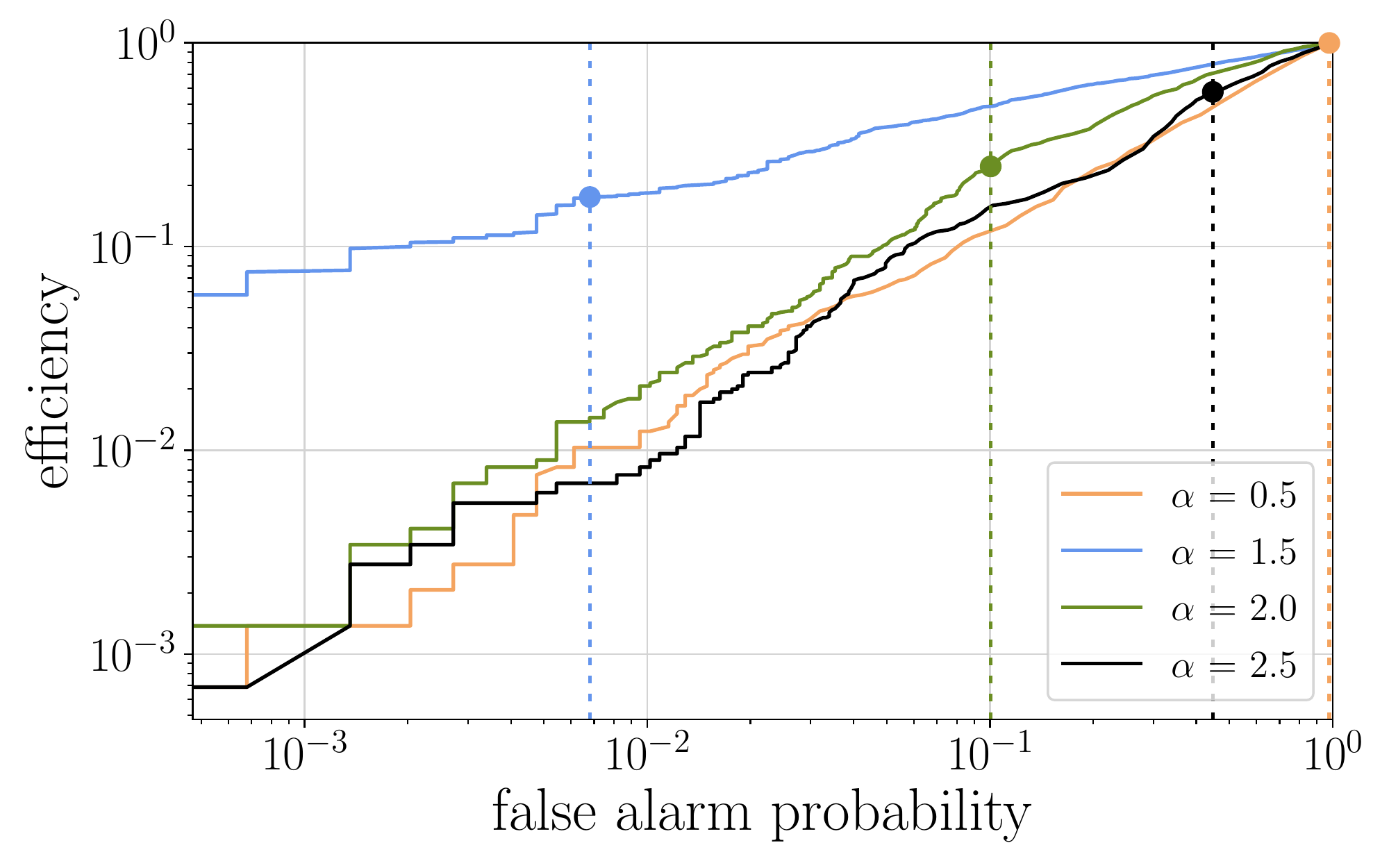}\hfil
\end{center}	
	\caption{\label{f:ROCm} Receiver Operating Characteristic curves for the detection of $m=1,3,4,5$ modes (orange, blue, green and black curves respectively) for the gravitational wave GW190814. 
The ROC curves 
for $\alpha=0.5$, $2.0$, and $2.5$ (i.e. for $m= 1$, 4, and 5) the are not very informative as these multipoles are quite weak in this case. Vertical lines and dots show the on-source \textit{p}-values with the same color code: the \textit{p}-value corresponding to $m=3$ is definitely much smaller than those corresponding to the other multipoles, with a detection efficiency that is nearly as high.}
\end{figure*}

\FloatBarrier

As explained in the previous section, from these empirical distributions we obtain the detection efficiency and the false alarm probability as functions of the residual energy $E_\mathrm{res}$, and this defines a parametric curve (the ROC curve) in the (false~alarm~probability, detection~efficiency) space. 
 
We produced ROC curves with many different parameter choices. We found a weak dependence of the Receiver Operating Characteristic on variations of the settings about our initially defined operating point ($df = 16$~Hz, $\delta \alpha = 0.15$, $\Delta t = 0.5$~s, $\delta t = 0$~s). Tested settings included $df = 8, 16, 32$~Hz, {$\delta \alpha=0.10,0.15,0.25,0.40$, $\Delta t=0.35,0.40,0.45,0.50,0.60$~s, and $\delta t$ from 0 to 0.08~s with step size $0.01$~s}. As an example, figure \ref{f:ROC} shows the set of ROC curves obtained by scanning over $\delta \alpha$ with fixed $df$, $\Delta t$, and $\delta t$.
The final settings from ROC optimization are $df=16$~Hz, $\delta \alpha = 0.1$, $\Delta t = 0.5$~s. We found that both $\delta t = 0$~s and $\delta t = 0.03$~s gave similar results and the final choice of $\delta t = 0.03$~s has been determined by the need to match the time duration of one pixel, to avoid the inclusion of significant contributions from merger and post-merger effects, while still preserving the optimal ROC (uppermost curve in figure \ref{f:ROCm}). {In the tests that follow, we fixed these settings assuming them to be optimal.}

Some figures of merit are easily extracted from the optimal ROC curve: according to predictions based on {\tt SEOBNRv4HM\_ROM}, the cWB detection of the $(3,3)$ multipole in GW190814 is expected to give an 18\% (40\%) detection efficiency with a false alarm probability less than 1\% (5\%). On the whole, the method is fairly efficient even with a quite low false alarm probability for the $(3,3)$ multipole. 

We also remark that the optimization of the time-frequency slice has been carried out using only off-source data and simulated signals, and this means that it does not pose any additional condition on the significance of the on-source results

After optimizing our method for the $(3,3)$ multipole we repeated the construction of the empirical distributions for the whole $\alpha$-range from 0.25 to 2.75 with steps of 0.05. In particular, we considered  the ROC curves corresponding to $\alpha = 0.5, 2, 2.5$: the comparison with the curve for $\alpha = 1.5$ is shown in Figure \ref{f:ROCm}. We see that no other HMs are detected in addition to the $(3,3)$ mode $\alpha = 1.5$
However, in general, a scan over a wide range of $\alpha$ values using a minimally-modeled search is well-motivated because in favorable cases it may detect  excess energy emitted in subdominant modes that are not included in the reference {\tt SEOBNRv4HM\_ROM} waveform model (as noted earlier, it includes only the $(2,1)$, $(3,3)$, $(4,4)$, and $(5,5)$ modes).

\begin{figure*}[h!]
\begin{center}
\includegraphics[clip,width=0.8\textwidth]{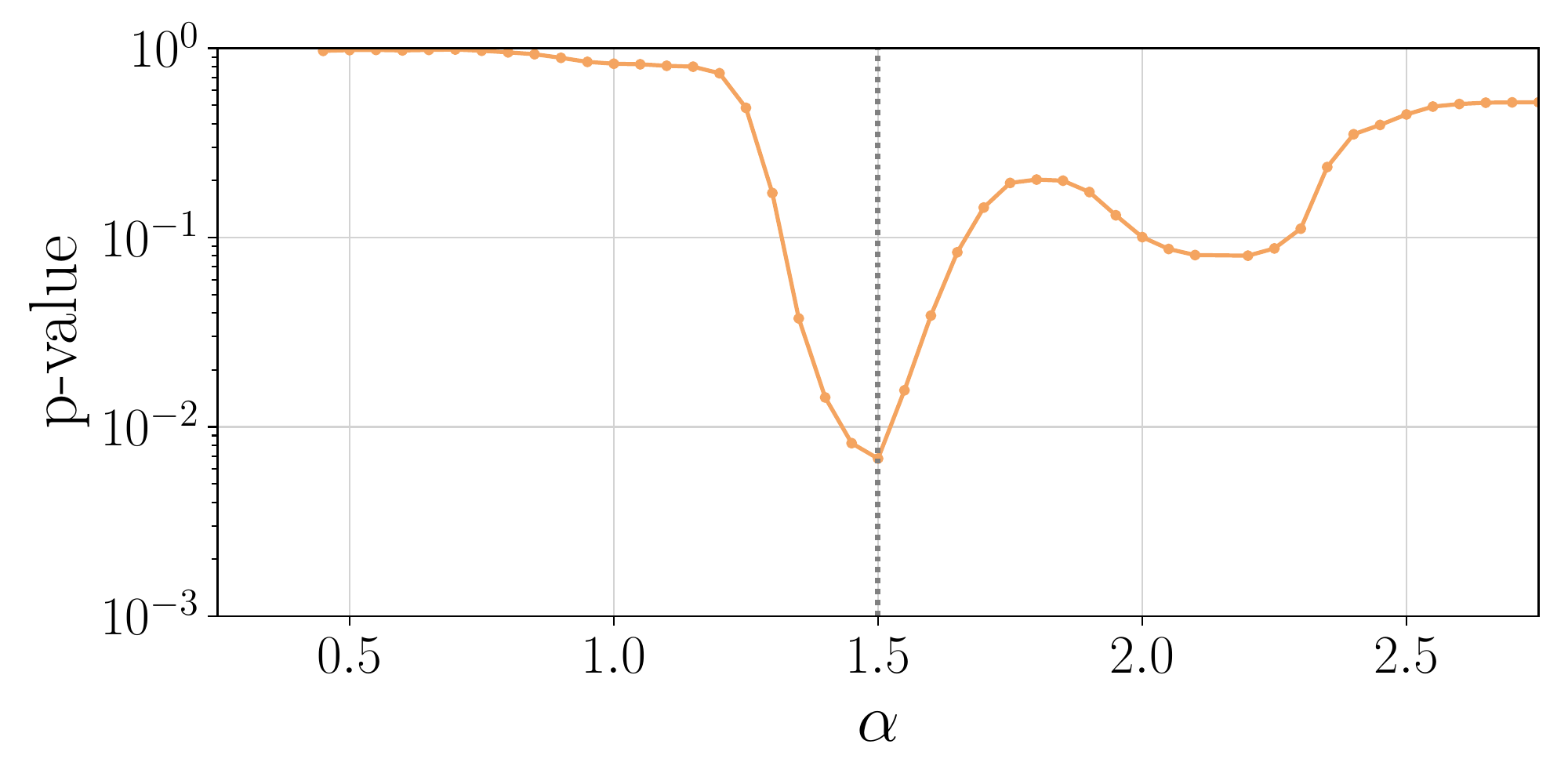}
\end{center}	
	\caption{\label{f:GW190814-pvalue} Plot of \textit{p}-value vs. $\alpha$ for the null hypothesis  (the {\tt SEOBNRv4\_ROM} waveform model), for the gravitational wave GW190814. The \textit{p}-value drops at a very low value at $\alpha =1.5$ mode, pointing to the presence of the corresponding $(3,3)$ multipole. We remark that the quadrupole fluctuation at $\alpha=1$ is compatible with the null hypothesis. {This figure is an enhanced version (extended $\alpha$-range and twice as many datapoints) of the lower panel of Fig. 7 in \cite{Abbott:2020khf}.}}
\end{figure*}

Figure \ref{f:GW190814-pvalue} shows the \textit{p}-value for the null hypothesis (the {\tt SEOBNRv4\_ROM} waveform model) as a function of $\alpha$ \cite{Abbott:2020khf}. 
The \textit{p}-value drops to a very low value at $\alpha =1.5$, pointing to the presence of the corresponding $(3,3)$ multipole, while the \textit{p}-value is much larger for other values of $\alpha$ and we could not reject the null hypothesis. 
We note in passing that cWB does not detect significant  residual energy at  $\alpha \sim 1$, which implies that the dominant $(2,2)$ multipole of the {\tt SEOBNRv4HM\_ROM} model is consistent with the {\tt SEOBNRv4\_ROM} waveform model.

\medskip

We end this section briefly considering the gravitational wave GW190412, another binary black hole system with asymmetric masses (we use the samples from Bayesian inference, and data from the LIGO Hanford, LIGO Livingston and Virgo interferometers that are publicly available at \cite{GW190412data}, the cWB reconstruction utilizes the public dataset available from \cite{gwosc}, see also \cite{O3details}). 

\begin{figure*}[ht!]
\includegraphics[clip,height=0.28\textheight]{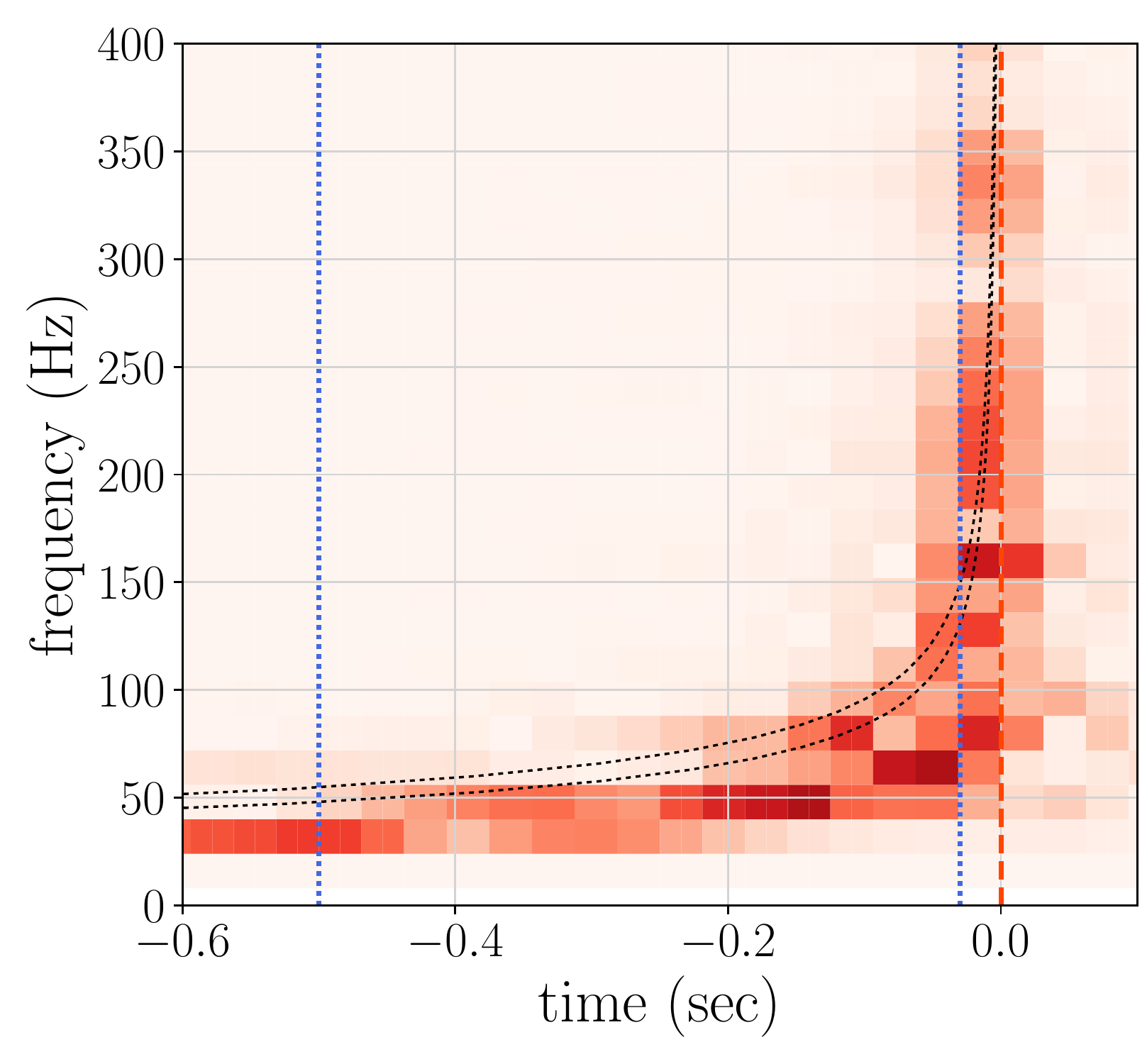} \hspace{0.02\textwidth}
\includegraphics[clip,height=0.28\textheight]{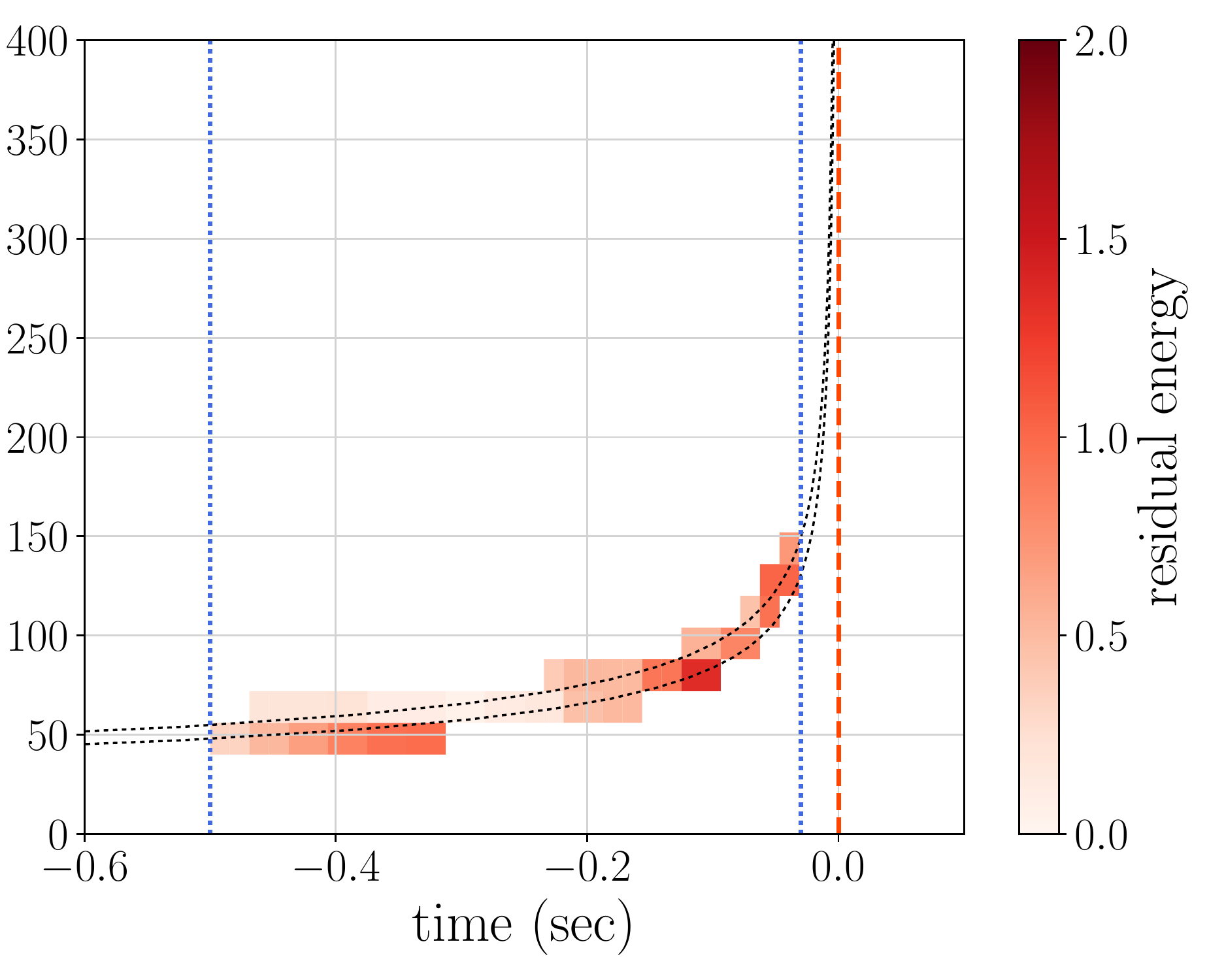} \\
\caption{\label{f:TFslice2} Left panel: time-frequency representation of the pixel-by-pixel residual energy of the GW190412 event obtained by cWB with respect to the MaxL {\tt SEOBNRv4\_ROM} waveform, projected into the LIGO Livingston detector using the WDM transform with resolution $dt = 1/32$~s and $df = 16$~Hz. Right panel: same map, showing only the time-frequency pixels which overlap at least partially the time-frequency slice with $\alpha=1.5$ (corresponding to the  $(3,3)$ HM): these pixels are used to evaluate the total residual energy $E_\mathrm{res}(\alpha; \delta \alpha, \Delta t, \delta t, df)$ in the time-frequency slice. The red vertical line shows the merger time  (GPS time: 1239082262.1658 s) from the MaxL {\tt SEOBNRv4\_ROM} waveform determined by Bayesian inference, referred to the LIGO Livingston detector. The dotted blue vertical lines show the selected time range of the time-frequency slice,  $[t_\mathrm{merger} - 0.5$~s$, t_\mathrm{merger} - 0.03$~s$]$. The black dotted curves show the boundaries of the time-frequency slice, defined by $[\alpha- 0.1, \alpha+ 0.1] \times f_{22}(t)$ with $\alpha = 1.5$. The fluctuations of $E_\mathrm{res}$ in pixels corresponding to the $(2,2)$ multipole in the left panel may appear large but they are compatible with the null hypothesis for this multipole (as shown later, see figure \ref{f:GW190814-pvalue}).}
\end{figure*}

\begin{figure*}[h!]
\begin{center}
\includegraphics[clip,width=0.8\textwidth]{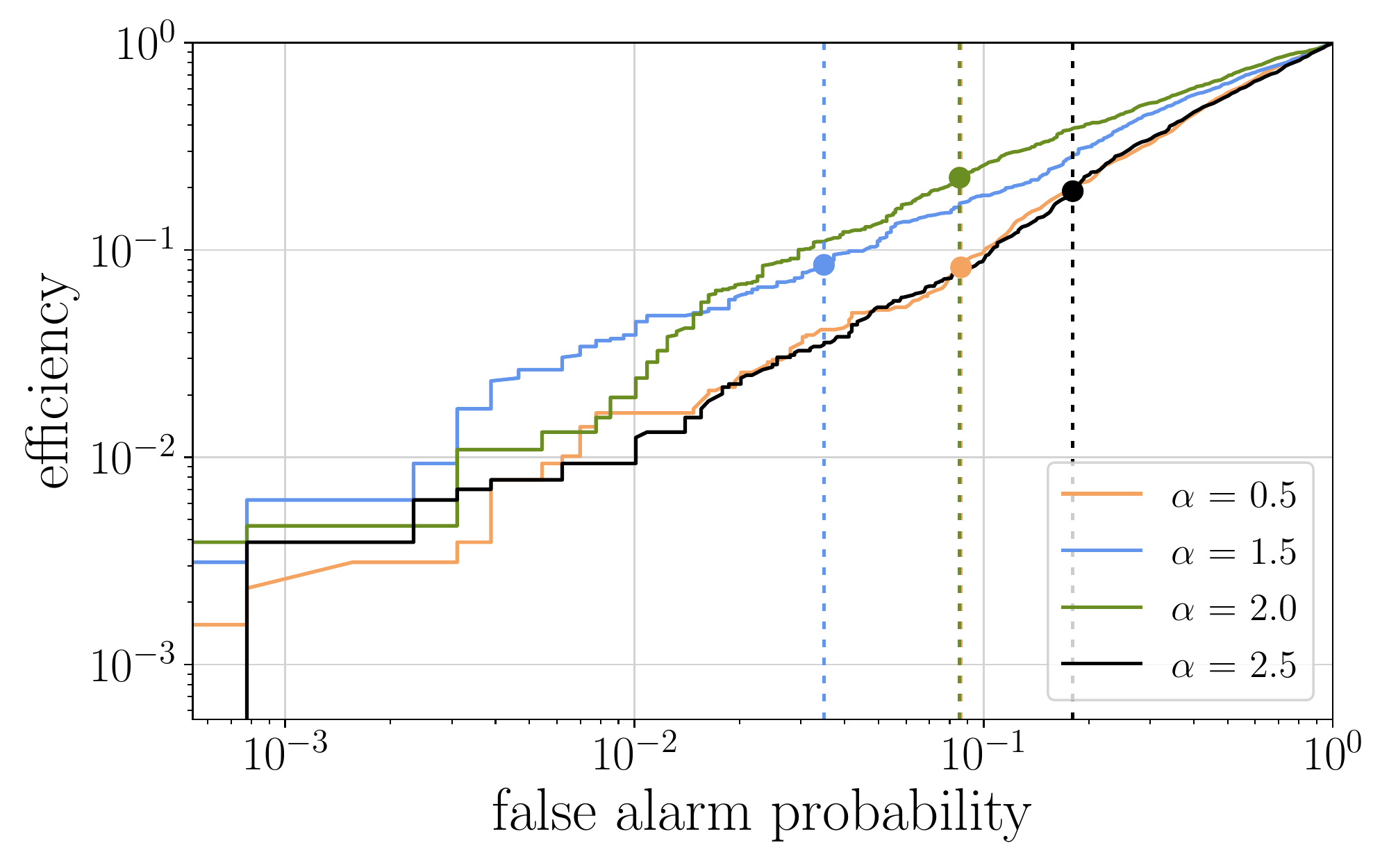}
\end{center}	
	\caption{\label{f:GW190412_roc} ROC curves for $E_\mathrm{res}(\alpha; \delta \alpha, \Delta t, \delta t, df)$, for the GW190412 gravitational wave, with the optimization found for GW190814. The best detection efficiencies are obtained for the $(3,3)$ and $(4,4)$ multipoles. }
\end{figure*}

\begin{figure*}[h!]
\begin{center}
\includegraphics[clip,width=0.8\textwidth]{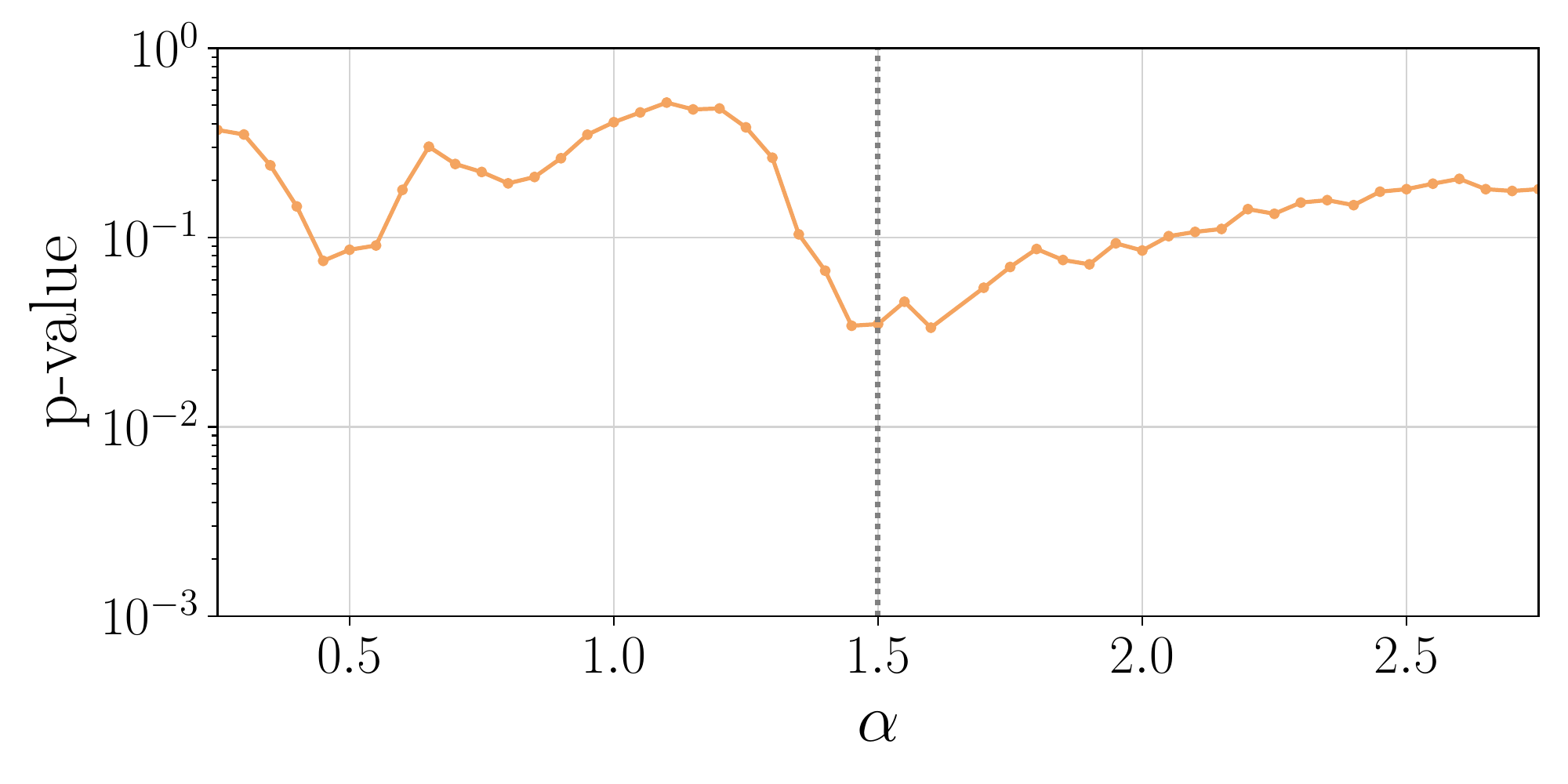}
\end{center}
	\caption{\label{f:GW190412-pvalue} Plot of the \textit{p}-value curve for the GW190412 gravitational wave, after optimization for the detection of the $(3,3)$ multipole. The plot shows a structure with a very shallow minimum at $\alpha=1.5$ with \textit{p}-value $< 0.03$: this hints at the need to introduce  higher multipoles at $\alpha = 1.5$. The dip at $\alpha=0.5$ with \textit{p}~value $\sim 0.08$ may also hint at the presence of the $(2,1)$ mode. However, \textit{p}~values such as these are common (as shown in figure \ref{f:nulltest}) and we cannot conclude anything in this case. }
\end{figure*}

We have repeated the procedure to optimize the slice parameters for $\alpha = 1.5$, and we have found  the same parameters
found for GW190814.  GW190412 has been produced by the coalescence of a black hole binary system with a significant mass unbalance \cite{LIGOScientific:2020stg}, and has been shown to display signs of higher multipoles. Still, the mass unbalance is smaller{, implying a smaller amplitude of the $(3,3)$ HM,} and the results for this gravitational wave are not as clear-cut as for GW190814. Figures \ref{f:TFslice2}, \ref{f:GW190412_roc}, and \ref{f:GW190412-pvalue} show the GW190412 equivalents of figures \ref{f:TFslice}, \ref{f:ROC}, and \ref{f:GW190814-pvalue}: in particular, the \textit{p}-value plot, Fig. \ref{f:GW190814-pvalue}, shows a structure with a very shallow minimum at $\alpha=1.5$ with \textit{p}-value $< 0.03$, and this can only hint at the presence of HMs at $\alpha = 1.5$ {(this is an enhanced version -- with an extended $\alpha$-range and twice as many datapoints --  of the lower panel of Fig. 7 in \cite{Abbott:2020khf}.)}.

\FloatBarrier

\subsection{Studies of \textit{p}-value curves}

\begin{figure*}[h!]
\begin{center}
\includegraphics[clip,width=0.8\textwidth]{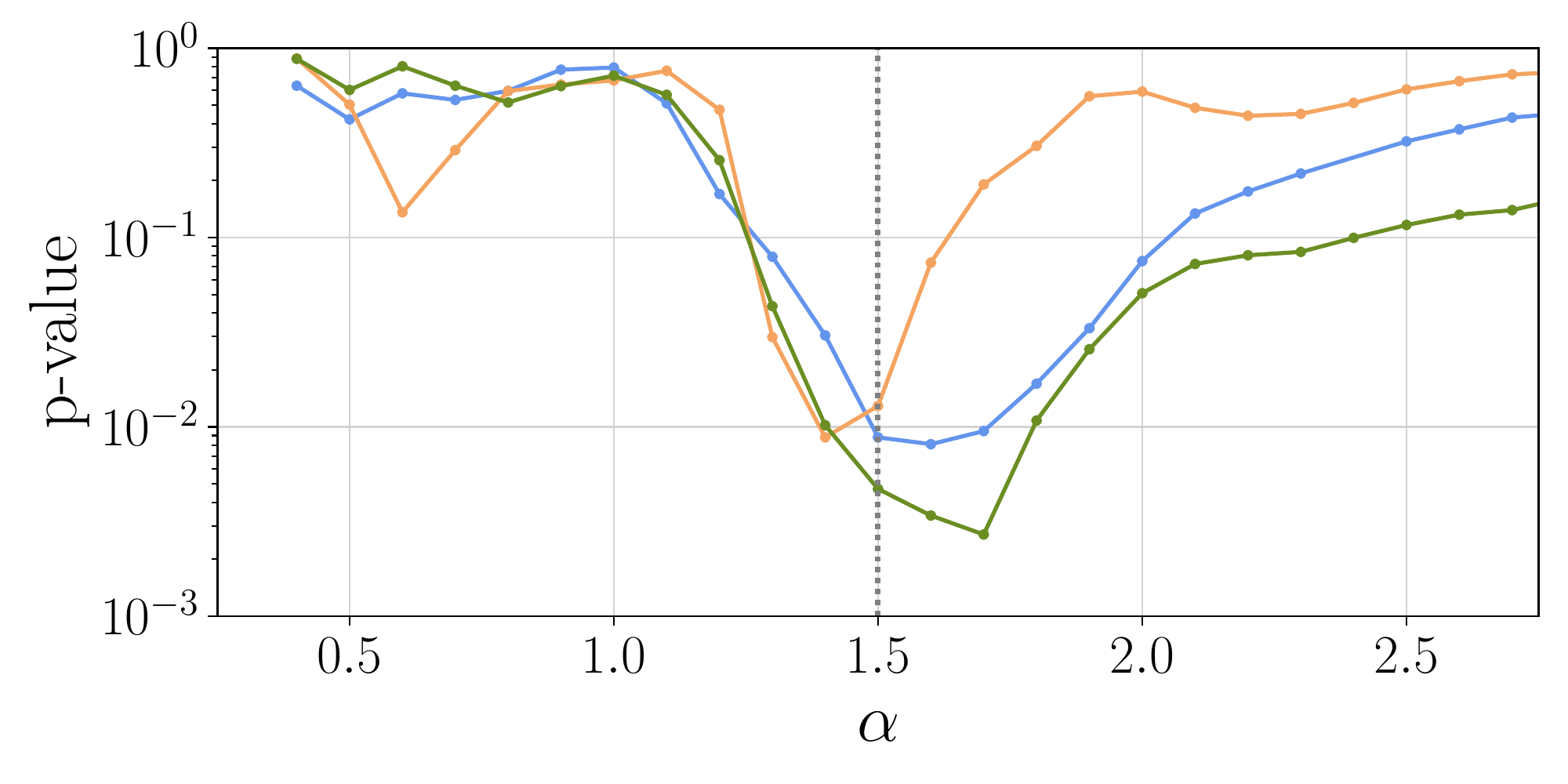}
\end{center}	
	\caption{\label{f:signaltest} Simulated \textit{p}-value curves for three different off-source injections of waveform posteriors from the {\tt SEOBNRv4HM\_ROM} waveform model for the gravitational wave GW190814. 
	The injections correspond to slightly different masses, inclinations, and signal amplitudes (blue curve: $m_1=24.52\;M_\odot$, $m_2=2.71\;M_\odot$, {$\iota = 54.49^\circ$,} SNR = 27.43; orange curve: $m_1=23.26\;M_\odot$, {$\iota = 132.14^\circ$,} $m_2=2.81\;M_\odot$, SNR = 25.39; green curve: $m_1=24.49\;M_\odot$, $m_2=2.71\;M_\odot$, {$\iota = 53.65^\circ$,} SNR = 23.85). These cases illustrate the  variety of  \textit{p}-value fluctuations.}
\end{figure*}

The behavior of the \textit{p}-value close to $\alpha = 1.5$ has been investigated for the event GW190814 using extended simulations to produce  \textit{p}-value curves like those shown in Fig. \ref{f:signaltest}. The three curves in the figure correspond to three specific waveform samples that have been selected from the larger injection set so that they satisfy the condition \textit{p}-value $\lesssim 0.01$ for  $\alpha = 1.5$. 
These \textit{p}-value curves are representative of the most interesting fraction (18\%) of the full set of results from injections in off-source data of waveform samples from the {\tt SEOBNRv4HM\_ROM} model.
Both the position of the \textit{p}-value minimum and the width of the dip vary according to the background noise encountered by cWB. 
We find that the \textit{p}-value curve shown in figure \ref{f:GW190814-pvalue} is qualitatively compatible with the fluctuations shown in figure \ref{f:signaltest}.

\begin{figure*}[h!]
\begin{center}
\includegraphics[clip,width=0.8\textwidth]{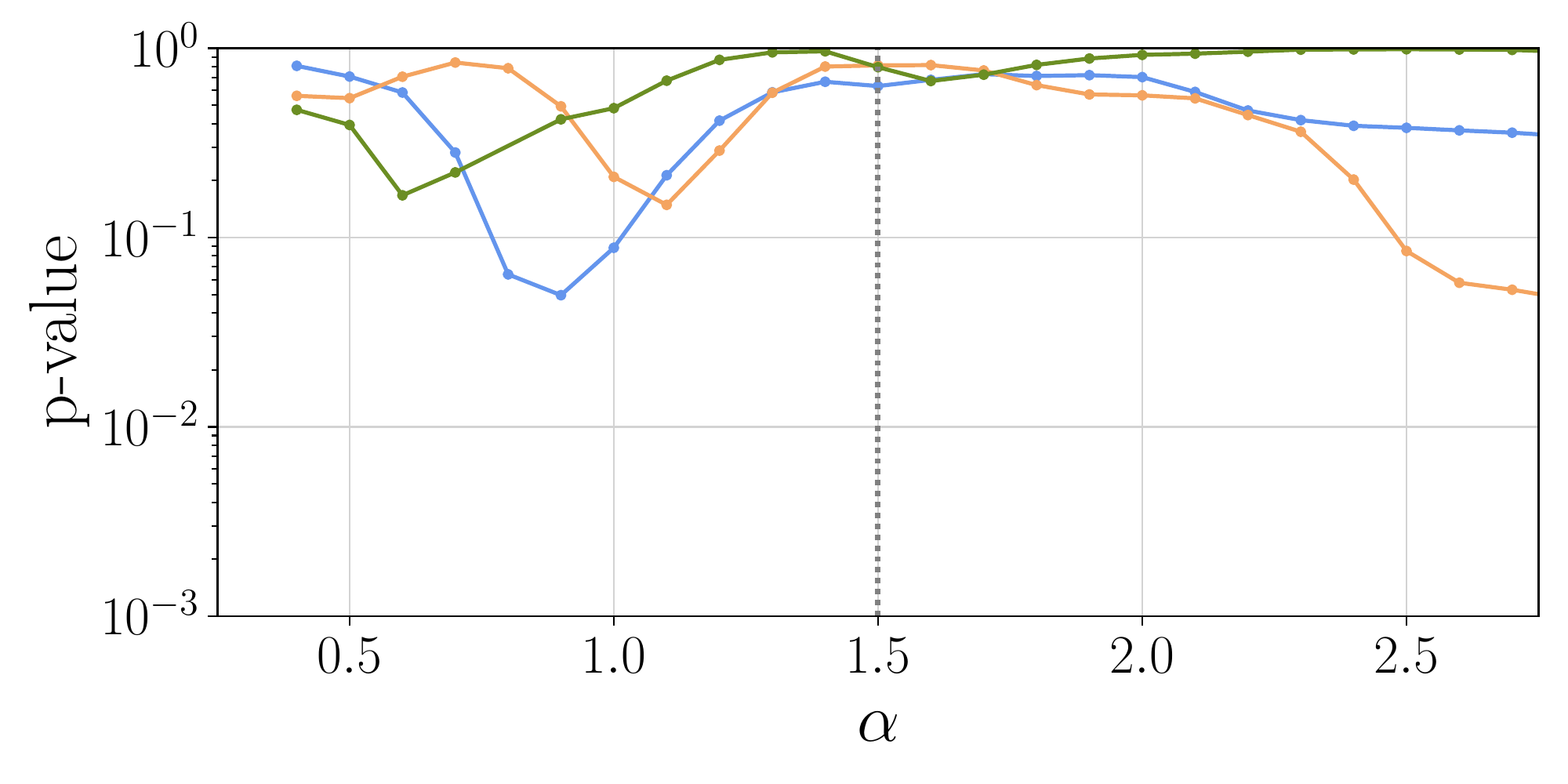}
\end{center}	
	\caption{\label{f:nulltest} Simulated \textit{p}-value curves from three additional off-source injections of waveform posteriors from the null signal model, {\tt SEOBNRv\_ROM}, performed within 30~s of GW190814. {In this case the waveform sample is always the same ($m_1=22.02\;M_\odot$, $m_2=2.90\;M_\odot$, $\iota = 14.58^\circ$) injected at different times (10~s, 20~s, and 30~s from the on-source event, returning different SNR's: blue curve SNR = 22.83; orange curve SNR = 22.78; green curve SNR = 23.03).}}
\end{figure*}

To further investigate the shape of the \textit{p}-value vs. $\alpha$ curves in the absence of higher multipoles, additional injections were performed off-source within 30~s of the GW190814 time using the {\tt SEOBNRv4\_ROM} model, and the results are shown in Figure \ref{f:nulltest}. 
In absence of HM emission, the \textit{p}-value vs. $\alpha$ curves show shallow dips driven by noise fluctuations with correlations for close $\alpha$ values.
This implies that in untargeted searches where we do not select a specific mode but scan a wide $\alpha$ region searching for minima at undefined $\alpha$ values, the corresponding \textit{p}-value must be corrected for the look-elsewhere-effect with a trials factor that takes this correlation into account. However, this is not the case in the current study where the main goal is to search for the more prominent fixed-$\alpha$ HM.

\section{Conclusions and final remarks}
\label{sec:remarks}

The method described here is an extension of our previous work \cite{PhysRevD.100.042003} and implements a new procedure to include robust a priori information on the specific feature of the gravitational-wave transient searched for: this is accomplished by focusing the coherent analysis of the data of the network of gravitational-wave detectors on a specific portion of the time-frequency representation of the signal, selected by optimizing the Receiver Operating Characteristic. The method is complementary with respect to that described in \cite{LIGOScientific:2020stg,Abbott:2020khf,Roy:2019phx}, however here we use the full detector network and its coherence, and we make no assumption of noise Gaussianity. There are also differences in the construction and numerical treatment of the time-frequency slices. 

We applied the method in the GW190814 discovery paper  \cite{Abbott:2020khf}, and the discussion in the present paper fills in all the details of that analysis. The gravitational wave GW190814 was emitted by a binary system with very asymmetric masses, and we detect the $(3,3)$ multipole emission by rejecting the null hypothesis (no HM) with \textit{p}-value =~0.68\%. Apart from the $(3,3)$ multipole, we cannot confirm any other significant discrepancy with respect to the dominant quadrupole emission in a wide range of possible undertones and overtones, from 0.25 to 2.75 times the main quadrupole emission frequency. 

Here, we repeat the analysis also in the case of the GW190412 event, but the evidence for the  $(3,3)$ multipole is  weaker in this case, even though there appears to be a dip in the \textit{p}-value curve for $\alpha=1.5$.

We wish to stress that the method is based on a consistency test between the waveform reconstructed by cWB, which makes minimal assumptions on the gravitational wave transient, and the parametric estimate from Bayesian inference, which is based on detailed waveform models. 
As a result, this study is peculiar in providing  a differential measurement of consistency between data -- as represented by the cWB reconstruction -- and two different waveform models. 
The final results can be expressed as frequentist \textit{p}-values or confidence intervals for the on-source measurement, assuming the correctness of the signal model.

The procedure can be extended to analyze other, different features of gravitational wave transients with well-modeled time-frequency representations, and such that a time-frequency region can be tailored to the scope. Applications include the investigation of spectral features like of post-merger emissions, precursors, and memory effects. Work is in progress to develop more of these capabilities and test them on actual observations.

\ack
\label{}

cWB makes use of data, software and/or web tools obtained from the Gravitational Wave Open Science Center \cite{gwosc}, a service of LIGO Laboratory, the LIGO Scientific Collaboration and the Virgo Collaboration. LIGO is funded by the U.S. National Science Foundation. Virgo is funded by the French Centre National de Recherche Scientifique (CNRS), the Italian Istituto Nazionale della Fisica Nucleare (INFN) and the Dutch Nikhef, with contributions by Polish and Hungarian institutes. 

The authors are grateful for computational resources provided by the LIGO Laboratory and supported by National Science Foundation Grants PHY-0757058 and PHY-0823459. 
ST is supported by University of Zurich Forschungskredit Nr. FK-19-114 and Swiss National Science Foundation. 
In addition, ST and AM acknowledge support from the European Gravitational Observatory, Convention EGO-DIR-63-2018.
\medskip 

{Some of the results discussed here appeared in preliminary form in the internal note \url{https://dcc.ligo.org/LIGO-T2000124/public}. }

\medskip

An open source release of cWB is available at \url{https://gitlab.com/gwburst/public/library} with related documentation at \url{https://gwburst.gitlab.io/documentation/latest/html/overview.html}.

\section*{References}
\bibliographystyle{iopart-num.bst}
\bibliography{paper}






\end{document}